\documentstyle[aps,pra,twocolumn,psfig,floats]{revtex}
\draft
\begin{document}

\title{ \vskip -0.5cm
         \hfill\hfil{\rm\normalsize Printed on \today}\\
         Nanoscale Processing by Adaptive Laser Pulses}

\author{Petr Kr\'al}

\address{Department of Chemical Physics, Weizmann Institute of Science,
         76100 Rehovot, Israel}

\maketitle

 
\begin{abstract}
We theoretically demonstrate that atomically-precise ``nanoscale processing" 
can be reproducibly performed by adaptive laser pulses. We present the new
approach on the controlled welding of crossed carbon nanotubes, giving various 
{\it metastable} junctions of interest. Adaptive laser pulses could be also 
used in preparation of other hybrid nanostructures.
\end{abstract}
 

 

\maketitle


Understanding and fully controlling growth, structural modifications and
coalescence of nanoscale materials has a top technological priority.
Many excellent examples of such systems can be found among the numerous 
types of recently discovered nanotubes \cite{Iijima91,Tenne,Chopra}. 
These unique materials 
are usually grown under rather poorly understood conditions using plasma 
\cite{PlasmaG}, laser \cite{LaserG} and other strongly non-equilibrium 
processes, where catalytic atoms can activate ``zipping" of chemical 
bonds \cite{Lee}. 

Although artificial nanoscale materials resemble bio-macromolecules, 
they do not possess ``inheritable memories", analogous to DNA, crucial 
for their exact reproduction. Thus our principal question is, if we can 
develop effective atomically-precise processing methods, that can 
reproducibly prepare such systems. This goal has been, for example, 
achieved in a molecular beam growth of superlattices, where nanoscale 
patterns in one dimension are practically under full control \cite{MBE}.

It would be especially attractive to achieve a precise control over some
advanced structural modifications. An example is coalescent ``welding" 
of tubular structures \cite{Coalescence}, which could lead in crossed 
nanotubes \cite{CrossNT} to preparation of strong light-weight nets.
As just demonstrated experimentally \cite{Terrones}, 
atomically-smooth welded nanotube structures can be induced by irradiation 
with electrons of MeV energies. Unfortunately, control over this process is 
so far limited, and thermal healing of the radiation-induced defects 
\cite{Ajayan} is only partial.

It thus becomes very interesting to consider {\it thermal} welding of 
nanotubes, driven by an externally applied pressure. Recent simulations 
\cite{Yakobson1,Yakobson2} have revealed that this approach could produce 
defect-free welded structures, with deep energy minima. On the other hand,
practical applications could also largely benefit from a reproducible 
preparation of various {\it metastable} atomically-precise structures, 
with shallow energy minima. An example are nanotube junctions with a 
``quasi-continuous" variation of their atomic structures. 

Here, we address this challenging task, and explore the possibility of a
reproducible welding of nanotubes, by ``adaptive" laser pulses. A feedback 
control of ultrashort chirped pulses is already applied in traditional 
macro-welding of thin films \cite{Layers}. A precise nanoscale processing 
could, in principle, be performed by more sophisticated pulses \cite{pulses}, 
prepared in {\it optimal control techniques} \cite{OCT}, that can 
selectively break molecular bonds in gases and liquids \cite{COHEXP}. 
Such pulses possess many degrees of freedom in their complex shaping 
and tuning, with the potential for a storage of ``production informations".  

\begin{figure}[t]
\vspace*{-40mm}
\centerline{\psfig{figure=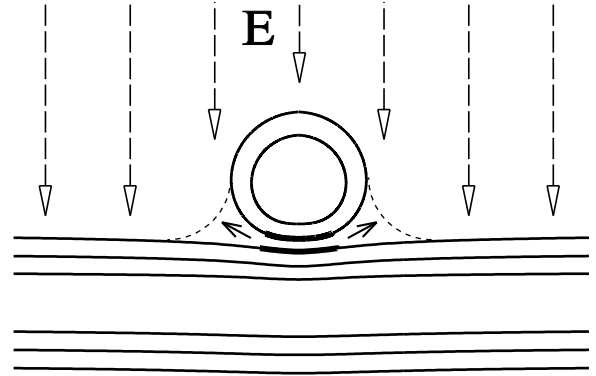,width=1.7\linewidth}}
\vspace*{-10mm}
\caption{Coalescent welding of crossed single-wall (SWNT) or multi-wall 
(MWNT) carbon nanotubes induced by adaptive laser pulses. Local time-dependent 
excitation heats a micron-size region of the nanotubes, and enables 
controlled atomically-precise restructuralization of C bonds at the 
pressured tube crossing (thin dashed lines).}
\label{WELD1}
\end{figure}

In Fig.~\ref{WELD1}, we present a scheme of the controlled welding of crossed 
nanotubes, realized by their local excitation with adaptive laser pulses. 
These pulses induce interband electron transitions in a micron-size region, 
where the hot carriers emit LO phonons. The generated phonons locally decay 
and heat nanotubes below their melting point ($T_{melt} \approx 4000$ K) 
\cite{MeltC}. Reconstruction of C bonds is thus induced {\it thermally}
in the contact area, where nucleation energies of potential defects are 
decreased by external pressure. Since the flipped bonds are selected 
and directed by the configuration of the whole system and the applied pressure, 
light pulses mostly control the total {\it extent} of coalescence. The 
joined region preserve the hexagonal atomic pattern if the tube structures are 
complementary (armchair/zigzag). 

We start the analysis of the welding control with the description of the 
nanotube excitation by inhomogeneous light intensity $E(x,t)$ 
\cite{klubka}. From the Fermi's Golden rule we can estimate the 
electron/hole injection rate, $\dot{n}_{e(h)}(x,t)\propto |E(x,t)|^2$, 
in the conduction/valence band of the crossed SWNT. Typically, we use the 
field intensity $E_0=5$ MV/cm at frequency $\omega= 10^{15}\, s^{-1}$, 
for which we obtain $\dot{n}_{e(h)}\approx 0.3$ ps$^{-1}$ per unit cell, 
with 40 C atoms for the (10,10) nanotube \cite{Mele}. This relatively strong 
excitation could cause variation of the optical coupling, and even lead to a 
dielectric breakdown of the system. Nevertheless, it should be possible 
to adapt the light pulses to these conditions in the learning process,  
so the structures could be controllably prepared without destruction.

The hot electrons and holes are generated with energies of a few 
($l\approx 3$) LO phonons, which are emitted within $\tau \approx 0.5$ ps. 
Since the carrier velocity is $v_e\approx 1\, \mu$m/ps, their 
energy is relaxed within the micron-size generation volume. Therefore,
the phonon injection rate $\dot{n}_p(x,t)$ approximately follows the 
excited electron/hole profile $\dot{n}_p(x,t)\approx l\, \bigl( \dot{n}_{e}
(x,t) +\dot{n}_{h}(x,t) \bigr)$. The LO phonons decay locally into LA phonons, 
which carry most of the heat. Thus the linear density of heat ${\cal H}(x,t)$, 
generated in each of the crossed nanotubes, is of the form
\begin{eqnarray}
{\cal H}(x,t) \approx {\cal E}_p\, \dot{n}_p(x,t)\propto 2\, l\,{\cal E}_p\,
 |E(x,t)|^2\, ,
\label{Hsou}
\end{eqnarray}
where ${\cal E}_p=h\nu_0$ is the LO phonon energy.

We can describe the one dimensional transport of heat by the equation 
\begin{eqnarray}
S\, \frac{\partial }{\partial x}
\Bigl( \kappa(T)\, \frac{\partial T}{\partial x} \Bigr)
+{\cal H}(x,t)=C(T)\, \rho\, \frac{\partial T}{\partial t}\ .
\label{Heat}
\end{eqnarray}
Here, $\kappa(T)$ is thermal conductivity, $C(T)$ specific heat, $S\approx 
1.18$ nm$^2$ the effective surface per tube in a rope and $\rho=1.9 \cdot 
10^{-23}$ kg/nm the mass density of the considered (10,10) armchair nanotube.
Both theoretical \cite{Berber} and experimental \cite{Hone} results in 
nanotubes lead to a large $\kappa(T)$, dominated by a phonon transport. 
From the predicted value $\kappa(T=300\, K) \approx 7500$ W/K$\cdot$m in 
SWNT, adjusted to the high-temperature profile in graphite 
\cite{DresselhausHT}, we obtain $\kappa(T)\approx 23.5\cdot 10^3/T^{1/5}$ 
W/K$\cdot$m. We can also use the specific heat of graphite 
\cite{DresselhausHT}, with the high-temperature fit $C(T)\approx 830+28.2\, 
(T-300)^{1/2}$ J/kg$\cdot$K. 

The thermally induced coalescence of nanotubes is 
driven by lessening of the pressure-assisted potential energy of 
the system (enthalpy), as in sintering. It can be realized 
by sequences of created and annihilated Stone-Wales (SW) 5/7 defects 
\cite{Yakobson2}, and some other types of defects \cite{Terrones}. Typically,
the energy of the SW defect is $E_{f}\approx 2 -\varepsilon\, C_0$ eV, 
where $\varepsilon$ is a local strain \cite{Yakobson1}. 
The formation of defects is limited by the size of their activation barrier 
$E^*\approx 6 -\varepsilon\, C_0^*$ eV, which can block the process even 
if $E_{f}< 0$. The last condition can be met already at strains $\varepsilon
\approx 0.1$, since $|C_0|\approx |C_0^*| \approx 20-30$ eV. Numerical 
simulations show \cite{Yakobson1} that the magnitude and sign of the 
material parameters $C_0$ and $C_0^*$ depend, in general, on the defect 
orientation and the tube chirality. Therefore, at the tube crossing, 
where the pressure-induced strain $\varepsilon$ is large, only defects 
with large and negative $C_0$ and $C_0^*$ would be thermally nucleated.

This fact could significantly reduce the large number of possible coalescence 
paths, switching between various intermediate structures. The nucleated 
defects are predominantly chosen by the types of nanotubes, geometry of their 
crossing and the applied pressure. Therefore, fluctuations of the followed 
coalescence paths should be relatively small, so that welded structures 
with the same number of defects could be reasonably {\it similar}. 
The most favorable path contains structures with the lowest energies. For 
example, a head-to-head coalescence of two (10,10) C nanotubes can be 
realized by following a path with just 68 steps of SW defects 
\cite{Yakobson2}. In this process, the flipped bonds first interconnect 
the tubes, soon after a neck is formed, and finally the tubes become 
smoothly rejoined (see Fig.~\ref{WELD1}). 
The goal of the optimal control is to {\it deliberately stop} this 
process at any of the metastable intermediate structures, formed 
in the vicinity of the most favorable path.

Consider that the system occupies with the probability $p_i$ any of 
the structures with $i$ defects, formed in the welding process. 
The probability $p_i$ is described by the rate equation
\begin{eqnarray}
\dot{p}_i=\frac{p_{i-1}}{\tau_{i-1}^+}+\frac{p_{i+1}}{\tau_{i+1}^-}
- p_{i}\left(\frac{1}{\tau_{i}^+}+\frac{1}{\tau_{i}^-}\right) \ ,
\label{DPT}
\end{eqnarray}
where the nucleation time $\tau_{i}^+$ ($\tau_{i}^-$) gives the transition 
rate for creation of $i+1$ ($i-1$) defects, if the system has $i$ defects. 
The nucleation processes are activated by the temperature at the tube 
crossing, so the times are
\begin{eqnarray}
(\tau_{i}^{\pm})^{-1} \approx 2\, N_{at}\, \nu_{0}\,
\exp{\left[- E_{\pm}^*(i)/k_{b} T\right]}\, .
\label{tnucl}
\end{eqnarray}
Here, $N_{at}\approx 20$ is the effective number of C atoms in the local 
region, $\nu_{0}=10$ ps$^{-1}$ is the vibrational frequency, used also in
${\cal E}_p$, and $E_\pm^*(i)$ are activation barriers of the $i+1$-th 
and $i-1$-th defects for structures with $i$ defects.

We model these barriers as follows,
\begin{eqnarray}
E_\pm^*(i)=6-\Delta \exp[\bigl(-|i-j_0|\pm 0.5\bigr)/D_i]\ {\rm eV}\, ,
\label{Enucl}
\end{eqnarray}
where we assume that the pressure-induced barrier shift, $\Delta>0$, is 
exponentially relaxed with the growing number $i$ of defects, since the 
pressure (motion of the tubes) cannot follow the fast (pulsed) coalescence. 
In numerical testing, we apply ns welding pulses, separated by $\mu$s 
periods. After the first pulse, the strain 
$\varepsilon$ can return to its initial value in all structures, irrespective 
of their number of created defects $j_0$.  In the next pulse, structures with 
different number of defects $j_0$ thus effectively start their evolution 
with the same barriers $E_\pm^*(i=j_0)=6-\Delta \exp[\pm 0.5/D_i]$.  Here, 
the shifts $\pm0.5$ reflect the pressure-induced asymmetry in transitions 
increasing/decreasing the number of defects by one, which is the driving 
force of the coalescence. 

Let us now discuss in more details practical realization of the welding
control.  In experiments, 
we can search the optimal field $E(t)$ in a feedback learning loop (see 
inset of Fig.~\ref{WELD4}), where we impose our welding requirements. 
In contrast to the MBE \cite{MBE}, where growth of individual monolayers 
is controlled {\it in situ}, here the light pulses need to be adjusted 
to the experimentally obtained structures. Mathematically, this search 
minimizes certain {\it functional}, which reflects the welding requirements 
\cite{OCT} and enables theoretical testing of the method. Since our goal 
is to reproducibly obtain certain metastable structures, {\it i.e.} 
structures with approximately the same given number ${\cal N}_D$ of 
defects, we look for a field $E(x,t)$ which minimizes the difference 
\begin{eqnarray}
F & = & |\langle {\cal N} \rangle- {\cal N}_D|\, , \ \ \ 
\langle {\cal N}^n \rangle=\sum_i\, p_i\, i^n\,  .
\label{F}
\end{eqnarray}
Here $\langle {\cal N}^n \rangle$ are moments of the distribution $p_i$.
The variance $\sigma_D=\sqrt{ \langle 
{\cal N}^2 \rangle - \langle {\cal N} \rangle^2}$ describes fluctuations 
of defects in the prepared structures.

In order to control the nanotube welding, we can take advantage of the fact 
that the pressure-induced lowering of the barriers $E_\pm^*(i)$ stops during 
the {\it fast} creation of the first $D_i$ defects. Thus, we use short 
light pulses, heating the system in such a way, that {\it less} than $D_i$ 
defects are created by each of them, so that fluctuations in $\sigma_D$ 
remain suppressed by the growing barriers $E_\pm^*(i)$. If we try to induce 
{\it more} than $D_i$ defects by a single pulse, then the tubes would have
to be largely heated, since the barriers are {\it not} lowered any more by 
pressure. Therefore, structural fluctuations would significantly grow and 
welding could spread without control in the whole region of the crossed tubes. 
This means that metastable structures with more than $D_i$ defects would
have to be prepared by  {\it several} pulses, each generating $\langle 
{\cal N} \rangle < D_i$ defects.

\begin{figure}[t]
\vspace*{-10mm}
\hspace*{3mm}
\centerline{\psfig{figure=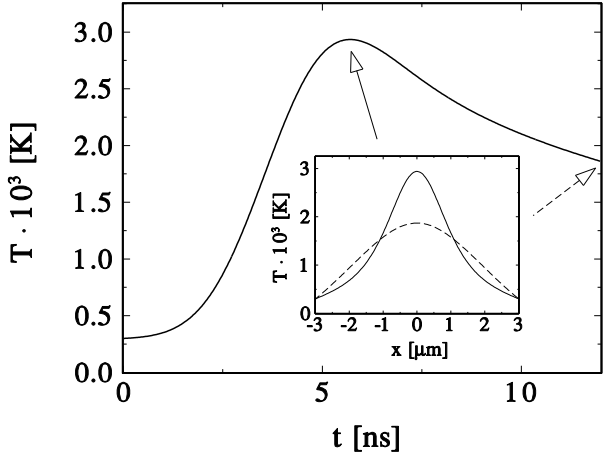,width=1.2\linewidth}}
\caption{Time-dependent temperature $T(x,t)$ at the crossing ($x=0$), 
induced by an optimal pulse. Three such (separated) pulses create in total
$\langle {\cal N} \rangle \approx 20$ defects. Spatial distributions 
of $T(x,t)$ at different times are shown in the inset. }
\label{WELD2}
\end{figure}

We now test these ideas by solving Eqns.~(\ref{Hsou}-\ref{F}).  Since the 
flipped bonds are largely selected by the system configuration (see above), 
the field practically only controls the {\it progress} of welding. 
Therefore, we can search it in a Gaussian form $E(x,t)=E_0\, 
\exp(-x^2/\sigma_x^2 -(t-t_0)^2/ \sigma_t^2)$, where we fix the parameters 
$\sigma_t =2.5$ ns ($t_0=4$ ns) and $\sigma_x =1\, \mu$m, and thus
choose $E_0$ as the only varied parameter.  In each iteration, of the 
learning scheme in Fig.~\ref{WELD4}, we apply three 
identical (separated) pulses to the system. Then we calculate the 
difference $F$, and let $E_0$ to evolve until $F$ is minimized. 
In real experiments, we can also vary the {\it shapes} of the pulses, in
order to fit more complex transient conditions. 

Here, we assume that the nanotubes are in a contact with each other at the 
crossing and with four contacts (two for each), separated $3\, \mu$m away 
from the tube crossing. They pull the two tubes in opposite directions 
and thus maintain the vertical force $F=5-15$ nN \cite{Yakobson2} at their 
crossing. We assume that the force leads to the model parameters $\Delta 
\approx 4$ eV and $D_i \approx 15$, used in Eqn.~(\ref{Enucl}).  The contacts 
also provide heat sinks, and are thus held at the temperature $T_0=300$ K.  
In this geometrical configuration, the nanotubes move and restore the force on 
$\mu$s timescale, after each welding pulse. 

\begin{figure}[t]
\vspace*{-10mm}
\centerline{\psfig{figure=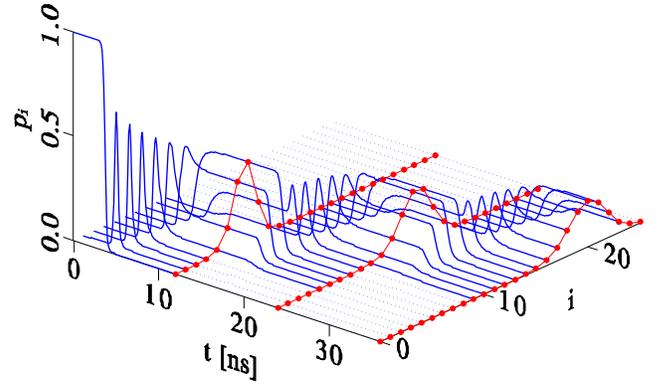,width=1.2\linewidth}}
\vspace*{-10mm}
\caption{Time-dependent probabilities $p_i(t)$ of structures with $i$ 
defects, obtained by excitation with three ns pulses, such as in 
Fig.~\ref{WELD2}. The pulses are separated by $\mu$s periods, denoted by
the dotted lines. With every pulse, the average number of defects moves 
up, until it reaches $\langle {\cal N} \rangle \approx 20$ defects.}
\label{WELD3}
\end{figure}

Figure \ref{WELD2} shows the temperatures $T(x=0,t)$ at the crossing, induced 
by excitation with any of the three (separated) laser pulses. By {\it 
requesting} that the pulses create in total $\langle {\cal N} \rangle \approx 
20$ defects, we have obtained their common optimal field amplitude, $E_0
\approx 5$ MV/cm, in several tens of iterations.  In the present pulsed 
regime the temperature rises up to $T \approx 3000$ K, 
then it slowly relaxes, as the heat diffuses through the tubes, and is 
pumped out through the contacts. In the inset, we show the related 
broadening and reshaping of the temperature profile.


In Fig.~\ref{WELD3}, we show the probabilities $p_i(t)$ of preparing 
structures with $i$ defects, by application of the above three pulses 
separated by $\mu$s time periods. The pressure-induced coalescence leads 
to narrow population maxima at certain numbers of defects $i$ after each 
pulse. The pulse creates less than $D_i$ defects, before the pressure is 
temporarily released and the coalescence stops. When the pressure rises 
again, the process continues with the next pulse, until $\langle {\cal N} 
\rangle \approx 20$ defects are created. 
In Fig.~\ref{WELD4}, we also present the time-dependence of the average 
number of defects $\langle {\cal N} \rangle$ and the variance $\sigma_D$,
for the excitation in Fig.~\ref{WELD3}. The simulation shows that $\langle 
{\cal N} \rangle$ largely depends on the transferred energy, while $\sigma_D$ 
remains small, if $\langle {\cal N} \rangle < D_1$ within one pulse.

\begin{figure}[t]
\vspace*{-10mm}
\hspace*{2mm}
\centerline{\psfig{figure=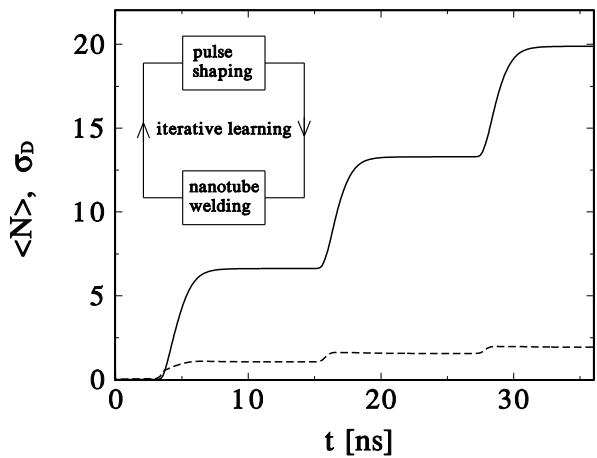,width=1.2\linewidth}}
\vspace*{-3mm}
\caption{Time-dependence of the average number of defects $\langle {\cal N} 
\rangle$ (solid line) and their variance $\sigma_D$ (dashed line) for 
the pulsed excitation in Fig.~\ref{WELD3}. In the inset, we show 
the learning scheme used in optimal control of the welding process.}
\label{WELD4}
\end{figure}

In order to speed up formation of nanotube nets, we can consider simultaneous 
irradiation of many such junctions under a {\it mask}, which would concentrate 
the light on the selected spots. Tubular nets might be also prepared by 
self-assembly processes, if we succeed to initiate nanotube branching, 
{\it i.e.} splitting of their growth in orthogonal directions.

With more sophisticated light pulses \cite{pulses,OCT,COHEXP} the 
method can also selectively catalyze chemical reactions 
inside or on the surfaces of nanotubes. This can lead to an efficient 
preparation of new hybrid materials, with {\it on-site} grown ``filling" 
and ``dressing". An example is light-induced polymerization or growth of 
nanocrystals inside nanotubes \cite{Sloan}. The adaptive pulses could even 
control electronic processes in nanodevices, especially if the chemical 
constituents can be reversibly switched by light \cite{Irie}. 

\vspace{3mm}
\noindent
I would like to acknowledge support from EU COCOMO. 


 \end{document}